# Oxygen distribution and segregation at grain boundaries in Nb and Ta-encapsulated Nb thin films for superconducting qubits

Jaeyel Lee[1*], Dieter Isheim[2,3*], Zuhawn Sung[1], Francesco Crisa[1], Sabrina Garattoni[1], Mustafa Bal[1], Cameron J. Kopas[4], Josh Y. Mutus[4], Hilal Cansizoglu[4], Jayss Marshall[4], Kameshwar Yadavalli[4], Dominic P. Goronzy[2], Mark C. Hersam[2,5,6], David N. Seidman[2,3], Alex Romanenko[1], Anna Grassellino[1], Akshay A. Murthy[1]

[1]Superconducting Quantum Materials and Systems Center (SQMS), Fermi National Accelerator Laboratory, Batavia, IL 60510, USA

[2]Department of Materials Science and Engineering, Northwestern University, Evanston, IL 60208

[3]Northwestern University Center for Atom Probe Tomography (NUCAPT), Evanston, IL 60208

[4]Rigetti Computing, Berkeley, CA 94710

[5]Department of Chemistry, Northwestern University, Evanston, IL 60208

[6]Department of Electrical and Computer Engineering, Northwestern University, Evanston, IL 60208

## Abstract

We report on atomic scale analyses of oxygen distribution and segregation at grain boundaries (GBs) of Nb and Ta-capped Nb (Ta/Nb) thin films for superconducting qubits using atom-probe tomography (APT) and transmission electron microscopy (TEM). We observe oxygen segregation at grain boundaries (GBs) relative to the oxygen concentration within the grains for both Nb and Ta-capped Nb thin films for superconducting qubits and find that higher oxygen concentration in the interior of Nb grains lead to greater oxygen segregation levels at GBs. This finding emphasizes that controlling oxygen impurities in Nb during film deposition and fabrication processing is important to reduce the level of oxygen segregation at GBs in Nb. The enrichment factor ($C_{GB}/C_{grain}$) for oxygen segregation at GBs in Nb is $2.7 \pm 0.3$ for Nb films, and Ta-capped Nb thin films exhibit slightly reduced Nb GB enrichment factors of $2.3 \pm 0.3$ while GBs in the Ta capping layer itself possess higher enrichment factors of $3.0 \pm 0.3$. We hypothesize that the Ta capping layer can trap oxygen and thereby affect oxygen in-diffusion and segregation at GBs in the underlying Nb thin films. Finally, we find that increases in the oxygen concentration in both Nb grains and GBs correlate with a suppression in the critical temperature for superconductivity ($T_c$). Together, our comparative chemical and charge transport property analyses provide atomic-scale insights into a potential mechanism contributing to decoherence in superconducting qubits.

*Equally contributed

## 1. Introduction

Superconducting qubits represent one of the most promising platforms for quantum computing hardware owing to fast gate operation, scalability, and compatibility with existing semiconductor manufacturing technologies [1-3]. Because the coherence time of superconducting qubits are limited by imperfections in materials introduced through the fabrication processes, understanding the relative impact of these material imperfections and developing mitigation strategies remain an active research area [4-12]. For instance, although Nb thin films have been widely employed as primary superconducting materials for superconducting qubits because of their relatively high critical temperatures and large superconducting gaps, the surface oxides on Nb thin films have been shown to act as a significant source of microwave loss and qubit decoherence [6, 13-16]. For this reason, significant efforts have been devoted to understanding the formation and properties of these oxides. Toward this end, recent materials advances, including the development of surface encapsulation of Nb films with other materials, such as Ta or Au, has led to systematic increases in the coherence times of superconducting qubits, which are attributed to the replacement of lossy $Nb_2O_5$ surface oxides with less lossy surface layers [13, 15, 17, 18].

In addition to surface oxides, the distribution of interstitial oxygen atoms within Nb thin films has been the subject of extensive investigations due to its potential role in pair-breaking and excitation of non-equilibrium quasiparticles [19-21]. In addition, oxygen impurities in Nb thin films have been shown to reduce superconducting gaps ($\Delta$) and suppress $T_c$, which also contribute to the excitation of quasiparticles [14, 21, 22]. Furthermore, magnetic impurities in superconductors are known to have a pronounced influence on superconducting properties [23-25]. In this context, the possibility that oxygen impurities in Nb thin films may act as localized magnetic moments and introduce microwave losses provides further motivation for characterizing the atomic-scale distribution of oxygen in the Nb thin films used in superconducting qubits [19, 26].

Similarly, oxygen segregation at grain boundaries (GBs) in Nb thin films has been proposed as a source of decoherence in superconducting qubits [14, 22]. Although oxygen enrichment at GBs in Nb thin films for superconducting qubits has been reported, previous studies have provided limited quantitative information regarding the concentration of oxygen impurities at GBs, as well as the factors impacting oxygen segregation at GBs [27, 28].

Toward this end, we employ atom-probe tomography (APT) combined with transmission electron microscopy (TEM) to analyze impurities in Nb grain interiors and at GBs to enable direct quantification of the atomic-scale distribution of oxygen within grains and at GBs [29-32]. Specifically, we investigate Nb and Ta-capped Nb thin films where the Ta-encapsulation prevents the formation of detrimental $Nb_2O_5$ on the surface of the film. Our studies reveal a strong correlation between the oxygen concentration in Nb grain interiors and the level of oxygen segregation at GBs, with enrichment factors ($C_{GB}/C_{grain}$) of 2.7 ± 0.3 for Nb thin films. Tantalum capping layers can trap oxygen and thereby potentially affect in-diffusion and segregation behavior of oxygen and the enrichment factors at GBs in Nb: with a Ta capping-layer a slightly decreased enrichment factor, 2.3 ± 0.3, was measured. The impact of oxygen impurities and GB segregation on the superconducting properties of Nb thin films are examined using physical property measurement systems (PPMS), which reveal a correlation between oxygen concentration and suppression of the superconducting critical temperature ($T_c$). Ultimately, this work provides unprecedented atomic-scale quantification of oxygen GB segregation, thus providing insight into one of the primary decoherence sources for superconducting qubits.

2. **Experimental procedures**

Nb and Ta-capped Nb thin films were deposited on Si (001) at Rigetti and on sapphire (0001) substrates at the Pritzker Nanofabrication Facility (PNF) at the University of Chicago. All films were deposited by DC magnetron sputtering. The Nb and Ta-capped Nb thin-film samples were characterized utilizing scanning transmission electron microscopy (STEM) and atom-probe tomography (APT). Sample preparations for TEM foils and APT nanotips were performed with a Thermo Fisher Helios 5 CX focused ion beam (FIB) at Fermi National Accelerator Laboratory (FNAL). The list of Nb and Ta-capped Nb thin-film samples investigated in the present study is provided in Table. 1. TEM foils and APT nanotips were initially thinned using $Ga^+$ ions at 30 kV with currents ranging from 24 pA to 0.79 nA, followed by low-energy milling at 5 kV with 15 pA to reduce surface radiation damage. STEM imaging of the Nb and Ta-capped Nb thin-film samples was performed using an aberration-corrected JEOL ARM 200. APT analyses were performed using a Cameca (Madison, WI) LEAP5000 XS. APT nanotips for Nb and Ta/Nb thin films were cooled to 40 K, and field evaporation was activated with a picosecond ultraviolet laser (wavelength = 355 nm) employing 30-45 pJ laser pulses at a pulse repetition rate of 500 kHz. The detection rate during data acquisition was maintained between 0.2-1.0 %. Three-dimensional (3-D) reconstructions of the nanotips were performed using IVAS 3.8 (Cameca, Madison, WI). Resistivities of the films were characterized using four-point charge transport measurements as a function of cooling temperature to determine the superconducting critical temperatures ($T_c$) of Nb thin-films in a Quantum Design (QD) 9T physical property measurement system (PPMS).

Table. 1 List of Nb and Ta-capped Nb (Ta/Nb) samples analyzed in the current study and summary of analyses of oxygen distributions and segregation at GBs in Nb and Ta capped Nb thin-films.

| No. | Materials | Substrate | Comment |
| --- | --- | --- | --- |
| Sample 1 | Nb | $Al_2O_3$ (0001) | Magnetron sputtering at RT |
| Sample 2 | Nb | Si (100) | Magnetron sputtering at RT |
| Sample 3 | Ta/Nb | $Al_2O_3$ (0001) | Magnetron sputtering at RT |
| Sample 4 | Ta/Nb | Si (100) | Magnetron sputtering at RT |
| Sample 5 | Ta/Nb | $Al_2O_3$ (0001) | Magnetron sputtering at RT / post-annealed at 800 ºC for 3 h |

## 3. Results

An annular dark-field (ADF)-STEM image of a 200 nm thick polycrystalline Nb thin-film on (0001) sapphire substrate (Sample 1) is presented in **Fig. 1(a)**. From this image, we observe the typical columnar grain structure of sputtered Nb thin films with 20-50 nm grain diameters. A 3D reconstruction of an APT nanotip prepared from Sample 1 is displayed in **Fig. 1(b)**, including the Ni protective layer coating at the top and the sapphire substrate at the bottom. A detailed APT 3D-reconstruction highlighting the distributions of Nb-O molecular ions in a nanotip of the Nb film (Sample 1) containing two GBs is displayed in **Fig. 1(c)**, demonstrating that oxygen is enriched along GBs of the columnar grains in the Nb thin film. Specifically, the oxygen concentration profile across the Nb surface oxide layer into the Nb bulk in **Fig. 1(d)** reveals that the oxygen concentration is highest near the surface of Nb (~10 at.%) and gradually decreases to $2.0 \pm 0.2$ at.% as the distance from the surface of Nb increases. An oxygen concentration profile across the two GBs, **Fig. 1(c),** measured 60 nm away from the Nb surface, demonstrates a local increase of the oxygen concentration at the GBs to $6.0 \pm 0.6$ at.% and $5.5 \pm 0.6$ at.% for GB1 and GB2, respectively, whereas the oxygen concentrations in the adjacent Nb grain interiors are 2.0 to $2.2 \pm 0.2$ at.%. Taking the average oxygen concentration of the Nb grain interior as $2.1 \pm 0.2$ at.%, the enrichment factor of oxygen at the GBs ($C_{GB}/C_{grain}$) are $2.9 \pm 0.3$ and $2.6 \pm 0.3$ for GB1 and GB2, respectively.

A similar analysis was performed for the Nb thin-film Sample 2 and the ADF-STEM image of Sample 2**, Fig. S1,** displaying a similar columnar microstructure with a 170 nm film thickness and 20-50 nm grain diameters. A 3D-reconstruction of the distribution of Nb-O molecular ions is presented in **Fig. S1(b)**. In this sample, a GB groove is observed at the surface of the Nb thin-film, and Nb-O molecular ions are enriched along the GB. The corresponding oxygen concentration profile across the GB in **Fig. S1(c)** displays a local increase of the oxygen concentration at the GB of $1.9 \pm 0.2$ at.%, whereas the oxygen concentrations in the adjacent Nb grain interiors are $0.5 \pm 0.1$ at.% and $0.7 \pm 0.1$ at.%, respectively. Taking the average oxygen concentration of the Nb grain interior as 0.6 at.%, the enrichment factor of oxygen at the GB ($C_{GB}/C_{grain}$) is $2.7 \pm 0.3$, which is similar to the values for Sample 1.

Despite the two Nb thin films displaying markedly different oxygen concentrations in the Nb grain interiors (i.e., 2.0 at.% O for Sample 1 and 0.6 at.% O for Sample 2), similar enrichment factors are measured for both samples. The difference in oxygen concentrations in Nb thin films is attributed to differences in the film deposition and fabrication conditions including the impurity levels of Nb in the sputter target, oxygen partial pressure in the sputtering chambers during the deposition process, and baking process for curing photoresist [20].

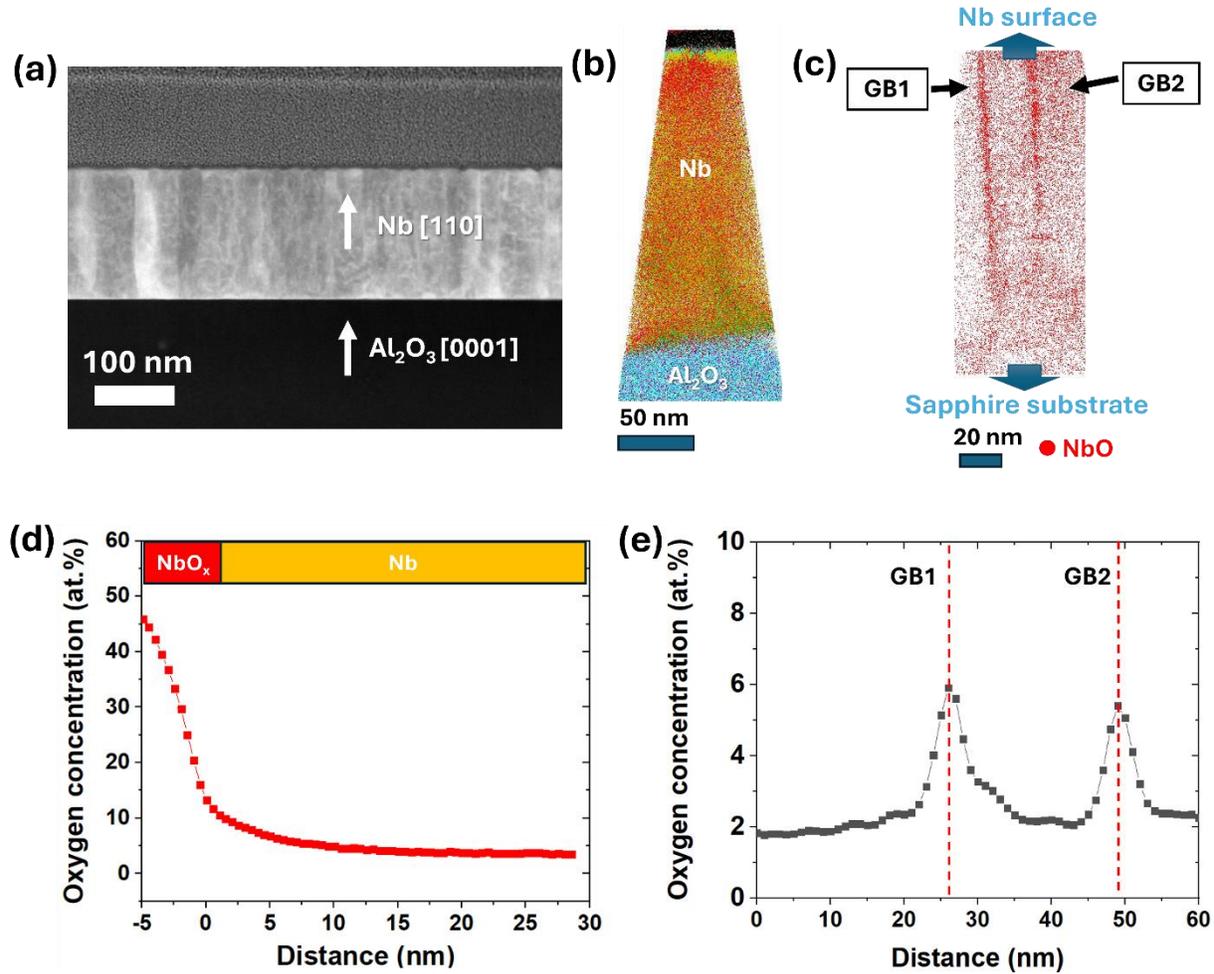

**Fig. 1** (a) Cross-sectional ADF-STEM image of a polycrystalline Nb thin film on a sapphire (0001) substrate (Sample 1) showing a 200 nm thick Nb film with 20-50 nm columnar grain diameters. (b) 3D APT reconstruction of polycrystalline Nb thin film on a sapphire substrate (Sample 1) is presented. (c) 3D reconstruction of Nb-O molecular ions for the Nb thin film (Sample 1) including two GBs is provided, and enrichment of Nb-O molecular ions at both GBs is observed. (d) An atomic concentration profile of oxygen atoms across the surface oxide (NbO$_x$)/Nb heterophase interface is obtained using the proximity histogram methodology [33], which reveals an oxygen concentration gradient in the film from the surface with high oxygen concentration toward the bottom with low oxygen concentration. (e) Atomic concentration profiles of oxygen across two GBs in Fig. 1(c) indicates that the oxygen concentrations at GBs increase to maximum values of 6.0 ± 0.6 at.% and 5.5 ± 0.6 at.% for GB1 and GB2, respectively, compared to 2.1 ± 0.2 at.% O for the Nb grain interiors.

For the APT reconstructions of Nb thin film Sample 1, additional analyses were performed along the depth direction at four different distances from the Nb surface: 25, 50, 75, and 100 nm, labeled as regions 1, 2, 3, and 4, respectively, **Fig. 2(a)**. Cross-sectional slices of Nb-O molecular ions at four depths are displayed in **Fig. 2(b)**, demonstrating that Nb-O molecular ions are segregated at GBs throughout the Nb film. The overlays of four one-dimensional (1-D) atomic concentration profiles across GBs indicated by blue arrow in **Fig. 2(b)**, are displayed in **Fig. 2(c)**. These profiles confirm that both the oxygen concentrations in the Nb grain interiors and the oxygen concentration at the GBs decrease progressively from 25 nm (region 1) to 100 nm depth (region 4). Both the maximum oxygen concentration at the GB and the oxygen concentration in the Nb grain interior gradually decrease with increasing distance from the Nb surface, **Fig. 2(d)**. We note consistent enrichment factors ($C_{GB}/C_{grain} \approx 2.6 \pm 0.3$) along this GB from region 1 to region 4.

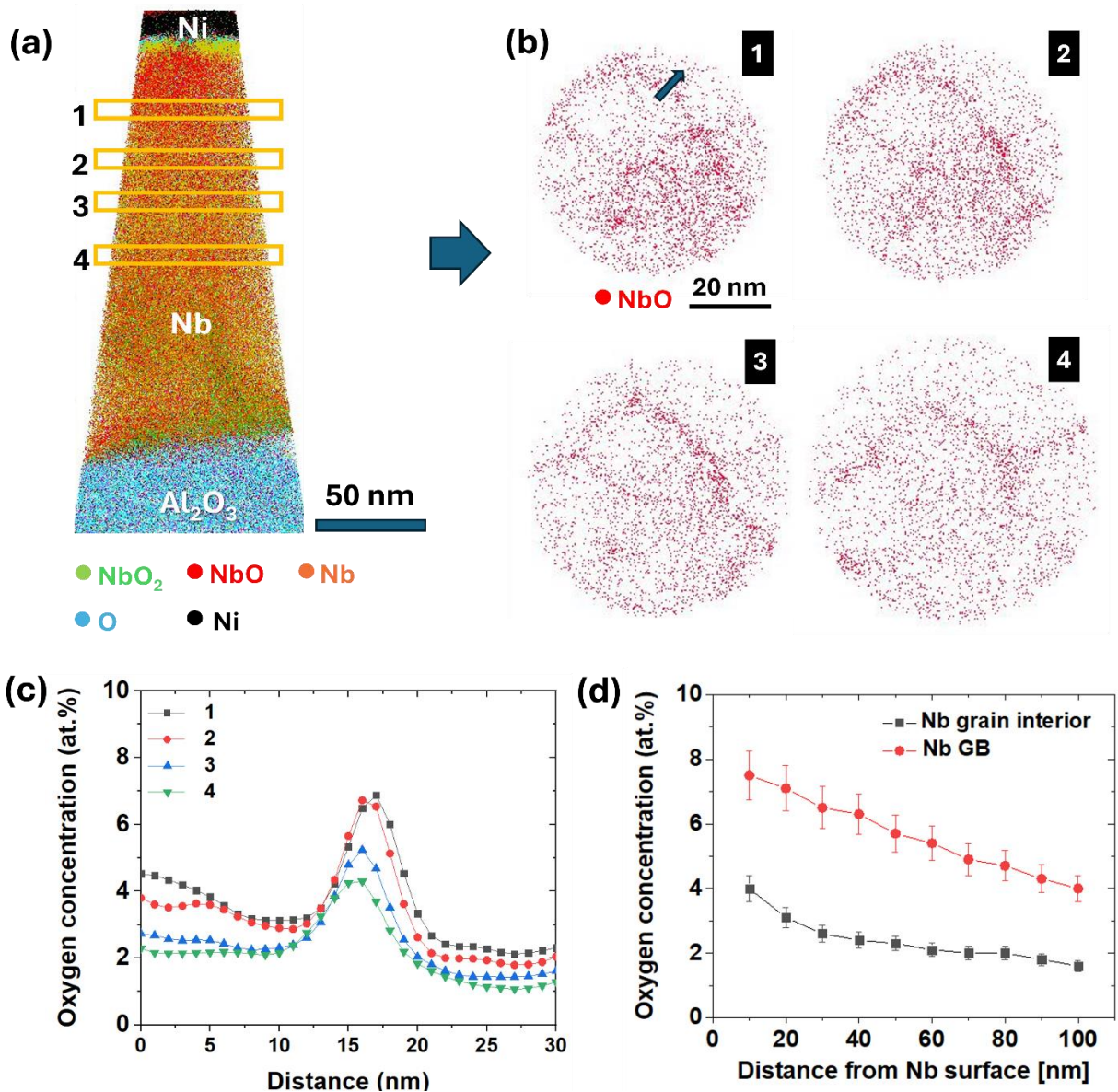

**Fig. 2** (a) 3-D APT reconstruction of an APT nanotip of a Nb thin film (Sample 1). (b) 3-D reconstructions of Nb-O molecular ions in four cross-sectional slices at different depths of the nanotip: (1) 25 nm; (2) 50 nm; (3) 75 nm; and (4) 100 nm; as indicated by the orange-colored boxes in (a). These depth-resolved reconstructions demonstrate that the oxygen concentrations in the Nb grain interiors and at the GBs decrease as the distance from the Nb surface increases. (c) Overlays of the four one-dimensional (1-D) atomic concentration profiles across the GB indicated by a blue arrow in (b-1) are presented. (d) Average oxygen concentrations at this GB and in neighboring Nb grain interiors are displayed as a function of distance from the Nb surface.

We next investigated Nb thin films capped with a 10 nm thick Ta layer (10 nm Ta/Nb) (Sample 3 and Sample 4) to examine the effect of a Ta capping layer on the oxygen distribution in Nb grain interiors and at Nb GBs. An ADF-STEM image of Ta-capped Nb (Sample 4) is presented in **Fig. 3(a)**, revealing surface grooves at GBs of the Ta capping layer on top of the undulating Nb thin-film. Note that body-centered cubic (b.c.c.) Ta (a = 3.3 Å) has the same crystal structure and lattice constant as b.c.c. Nb (a = 3.3 Å), permitting Ta capping layers to grow epitaxially on Nb [13]. APT analyses further confirm oxygen segregation at GBs in the Ta-capped Nb thin film. A 3D reconstruction of the APT nanotip including GB of a Ta-capped Nb on the surface of Ta/Nb thin film is presented in **Fig. 3(b)**, which exhibits an undulating groove on the surface of the Ta-capped Nb film similar to the one observed in the ADF-STEM images of the Nb thin-film. Both Ta-O and Nb-O molecular ions are enriched at the GBs, confirming that the undulating groove in the 3D reconstruction of Ta-capped Nb is on top of a GB with oxygen segregation. **Figures 3(c)** and **3(d)** display 3-D reconstructions with the positions of Ta-O and Nb-O molecular ions at the Ta/Nb interface along the in-plane direction. These reconstructions demonstrate that oxygen segregates not only at GBs in the Ta capping layer and but also continues along the GBs of the underlying Nb thin-film.

The 1-D concentration profiles of oxygen across the GB in the Ta capping layer and underlying Nb thin-film are displayed in **Fig. 3(e)** along with representative profiles across the Ta/Nb heterophase interface denoted by 1, 2, 3, 4, and 5. The O concentration in the Ta capping layer located at 1 shows 1.9 ± 0.2 at.% O on average, and the O concentration at the GB in the Ta layer is 5.7 ± 0.6 at.% O, which yields an enrichment factor ($C_{GB}/C_{grain}$) of 3.0 ± 0.3 slightly greater than the values for O segregation at GBs in Nb. For the Nb thin film below the Ta capping layer located at 5, the concentration is 0.6 ± 0.1 at.% O for the Nb grain interiors and 1.3 ± 0.1 at.% O at the GB, corresponding to an enrichment factor ($C_{GB}/C_{grain}$) of about 2.2 ± 0.2, which is slightly lower than the values for bare Nb thin films (i.e., 2.7 ± 0.3). Enrichment factors for O segregation at GBs at different distances from the Ta/Nb heterophase interface are summarized in **Fig. 3(f)**. APT analyses of GBs in Ta-capped Nb sample 3 were also performed, which exhibit a similar trend, **Fig. S2**.

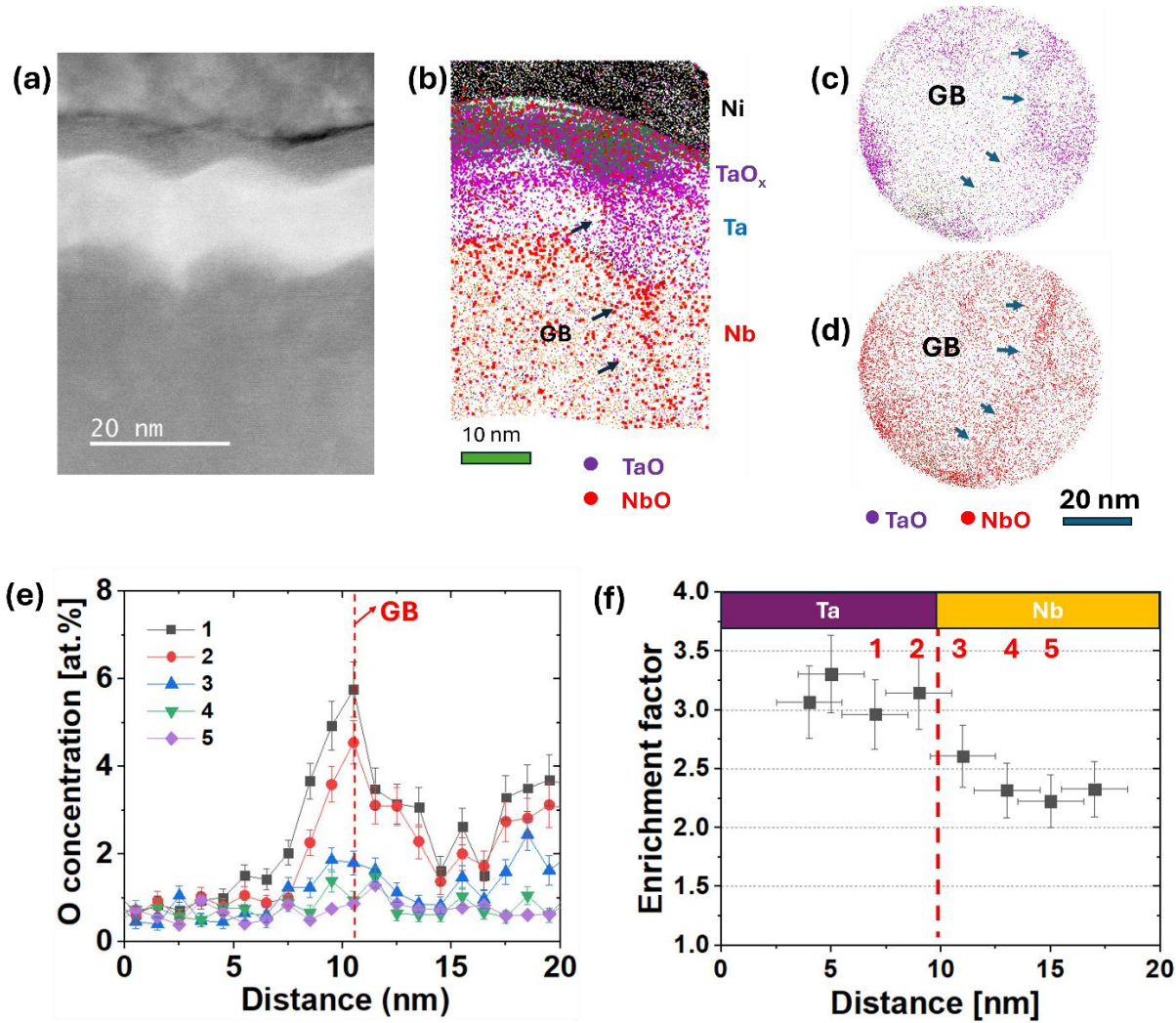

**Fig. 3** (a) ADF-STEM image of Ta-capped Nb Sample 4 exhibit an undulating surface with grooves at GBs of the Ta capping layer on top of the Nb thin film. (b) 3D reconstruction of an APT nanotip including a GB running through Ta and Nb layers is provided. It also illustrates an undulating surface and GB groove on the surface of the Ta-capped Nb film similar to the features in the ADF-STEM images of the Ta-capped Nb thin film, indicating that the groove in the 3D-reconstruction is a GB. 3D reconstructions of distributions of (c) Ta-O and (d) Nb-O molecular ions at the Ta/Nb interface in a horizontal direction are displayed. They confirm that Ta-O and Nb-O molecular ions are enriched at the GB of Ta-capped Nb thin film. (e) 1D concentration profiles of O across the GB in the Ta capping layer and Nb across the Ta/Nb heterophase interface located at 1, 2, 3, 4, and 5. (f) Enrichment factors for oxygen segregation at GBs ($C_{GB}/C_{grain}$) at different distances across the Ta/Nb heterophase interface are displayed.

**Figure 4 (a)** summarizes the O concentrations measured at GBs and within grain interiors for Nb, Nb with Ta-cap, and the Ta-capping layers. The data reveals a strong correlation between the oxygen concentration in grain interiors and that at GBs: higher O concentrations in interior Nb grains lead to higher oxygen concentrations at GBs in Nb thin films. Interfacial excess oxygen atoms at GBs are estimated for Nb, Nb capped with Ta, Ta capping layers, revealing approximately linear relationships between O concentrations in grain interiors and at GBs, **Fig. 4(b).** The enrichment factors ($C_{GB}/C_{grain}$) of each GB in the Nb and Ta-capped Nb thin-films, **Fig. 5**, demonstrate that the enrichment factors are 2.7 ± 0.3 for Nb thin-films and 2.3 ± 0.3 for Ta-capped Nb thin films, indicating that the presence of the Ta capping layer may affect oxygen segregation at GBs. Tantalum capping layers have larger O concentrations than the underlying Nb layers across Ta/Nb heterophase interfaces (**Figs. S2** and **S3**).

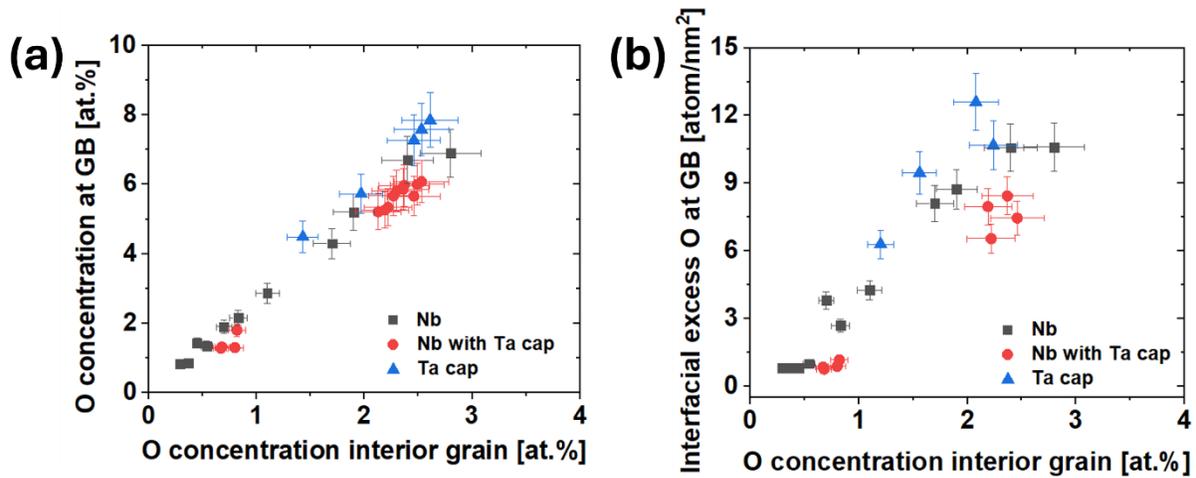

**Fig. 4** (a) Plot of O concentration at GBs (at.%) versus O concentration in grain interiors for Nb thin films, Ta-capped Nb thin films, and Ta capping layers is provided. It reveals a strong correlation between them. (b) Plot of interfacial excess of O at GBs (atom/square nm) versus O concentration (at.%) in Nb grain interiors.

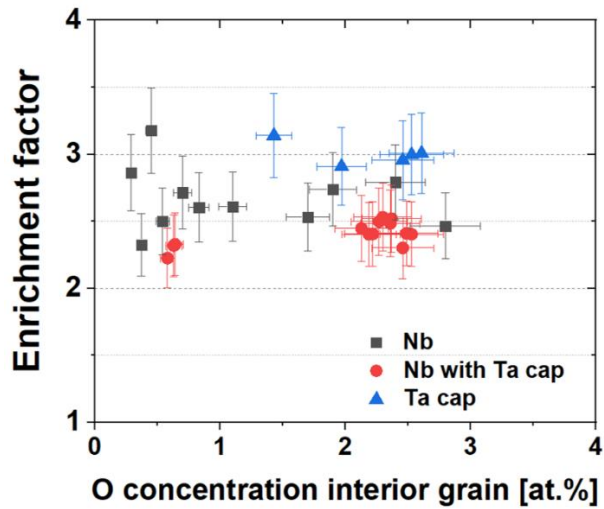

**Fig. 5** Summary of enrichment factors ($C_{GB}/C_{grain}$) for O segregation at GBs. These results show that the enrichment factor is reasonably consistent around 2.7 ± 0.3 for Nb thin films. We note that GBs in Ta-capped Nb thin-films show slightly smaller enrichment factors, 2.3 ± 0.3, and GBs in the Ta capping layers show slightly higher enrichment factors of 3.0±0.3.

Resistivity measurements were performed on selected Nb thin films using a Physical Property Measurement System (PPMS) to investigate the effects of O impurities on superconducting properties. The resistivity of Nb film Sample 1 around $T_c$ is a factor of three higher than Nb film Sample 2. This increased resistivity reflects the difference in O concentrations between the two films, **Fig. 1** and **S1**. The results demonstrate that the critical temperature ($T_c$) of Nb thin films gradually decreases with increasing O concentration in the Nb thin films, **Fig. 6**. Specifically, the Nb thin film Sample 1 containing 2.0 ± 0.4 at.% O in Nb grain interiors exhibits a $T_c$ of 8.9 K, whereas the Nb thin film Sample 2 with 0.5 ± 0.1 at.% O in Nb grain interiors has a $T_c$ of 9.3 K. This decrease in the Nb $T_c$ is smaller than earlier reports, which showed a reduction of 0.9 K in $T_c$ per 1 at.% O in Nb [19, 21]. This discrepancy may be attributed to inhomogeneities of the O distributions in the Nb thin-films, whereby regions with the smallest oxygen concentration in the Nb thin-films dominate the measured $T_c$ values. Additionally, it is possible that the effect of O on $T_c$ of Nb was overestimated in earlier studies [19, 21], due to the limitations in the experimental methods used for nanoscale quantification of oxygen impurities.

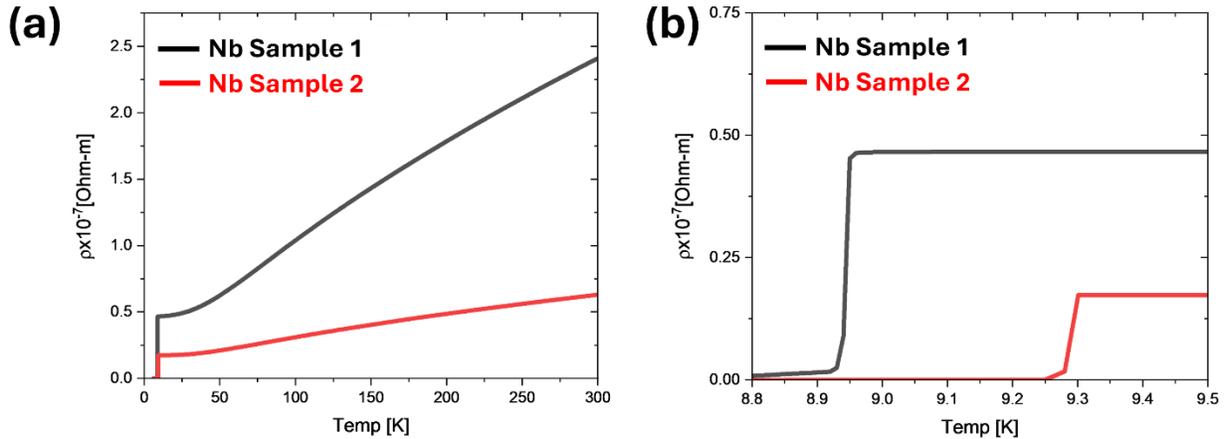

**Fig. 6** Resistivity measurements of Nb thin films for Sample 1 and Sample 2 using PPMS. Nb thin film Sample 1 yields a $T_c$ of 8.9 K due to the relatively large O concentration in the Nb thin film, and Nb thin film Sample 2 has a $T_c$ of 9.3 K which is near to the $T_c$ of pure Nb. The values in (a) and (b) present the onset of the superconducting transition temperatures, determined with the initial deviation from the normal conducting state.

## 4. Discussion

The total segregated oxygen excess at a grain boundary consists of three components: (i) a local equilibrium between the grain boundary and the oxygen bulk concentration in the locally adjacent grain; (ii) in-diffusion of oxygen supplied from the top surface of the Nb film; (iii) structures of GBs. We observe a strong correlation between the oxygen concentrations in Nb grain interiors and at GBs: a higher oxygen concentration in a Nb grain interior results in a higher oxygen concentration at GBs in Nb thin films, **Fig. 4(a)**. It demonstrates that local equilibrium between the GB and bulk oxygen concentration is the main contribution to determine the amount of oxygen segregation at GBs compared to the other two factors. According to the Nb-O phase diagram, the solubility limit of oxygen in Nb is about ~1 at.% at 600 °C in equilibrium with the NbO phase and it decreases to <0.1 at.% at room temperature (RT) [34, 35]. An excess interstitial oxygen atom concentration in Nb with respect to the solubility limit of oxygen in Nb can in principle segregate at GBs to minimize the free energy of the b.c.c. Nb phase if the formation of one of the niobium-oxide phases is kinetically delayed or suppressed. For Nb thin-film Sample 1, a higher baking temperature was used to cure the photoresist (200 °C for 35 min) compared to Sample 2 (120 °C for 60 min), which most likely promotes a greater oxygen incorporation due to the dissolution of the surface oxide into the bulk Nb [36, 37]. The Ta-capped Nb thin-film, Sample 5, was post-annealed at 800 °C for 3 h, which results in an oxygen concentration of >2 at.% O in the underlying Nb thin-film, **Fig. S3**. These results demonstrate that careful control of the deposition parameters, including the deposition and baking temperatures as well as the vacuum conditions, are essential for producing high-purity Nb thin films and for minimizing oxygen segregation at GBs.

In addition to segregation of bulk-dissolved oxygen at grain boundaries, another factor influencing the oxygen concentration at GBs in Nb thin films is diffusion of oxygen from the surface along GBs. 3-D reconstruction of APT nanotips, including GBs in **Fig. 1** and **Fig. 2,** illustrate an oxygen concentration gradient from the film surface with high oxygen concentration toward the bulk Nb, which has a small oxygen concentration. Because GBs are short-circuit diffusion paths, oxygen transport along GBs can contribute to the observed oxygen segregation at GBs additionally, **Fig. 1(c)** [38]. The Ta-capped Nb thin-films display a slightly smaller oxygen enrichment factor (2.3 ± 0.3) compared to Nb films (2.7 ± 0.3) with a $Nb_2O_5$ native oxide surface layer, **Fig. 5**. Tantalum has a slightly more negative heat of formation energy for $Ta_2O_5$ (2046 kJ/mole) than does Nb for $Nb_2O_5$ (1899 kJ/mole), which may lead to oxygen impurities becoming trapped in the Ta layer instead of diffusing to the underlying Nb [39, 40]. The Ta capping-layers contain a higher oxygen concentration than the underlying Nb thin films, **Figs. 3(e) and S2(b)**, suggesting that they possibly serve as diffusion barrier layers, thereby slowing or suppressing oxygen transport from the free surfaces [41]. This effect possibly reduces the supply of oxygen atoms for diffusion along GBs in the Nb film. Lastly, atomic structures of a specific type of GB affect the degree of segregation at GBs [42-44]. We observe some variation of oxygen concentrations at GBs even when the oxygen concentrations of adjacent Nb grain interiors are similar, **Fig. 1(e)**, which we attribute to possible differences in the atomic structures of each individual GB.

PPMS measurements confirm that increasing oxygen concentration in Nb thin films leads to a reduction of the critical temperature for superconductivity ($T_c$). With respect to the reduction of $T_c$ of Nb thin films with increasing oxygen concentrations, oxygen impurities and oxygen segregation at GBs in Nb thin films can contribute to the decoherence of superconducting qubits by introducing non-equilibrium quasiparticles in addition to other microscopic defects [12, 45, 46]. Microwave measurements of superconducting qubits based on Nb and Ta-capped Nb thin films investigated in the current study indicates that dielectric losses originating from surface oxides remain the dominant mechanism of decoherence relative to oxygen impurities and oxygen segregation at GBs [13]. Nevertheless, if the oxygen impurities in Nb and Ta/Nb thin films approaches a critical level, for instance, ~4 at.% O on average, then the contribution of excitation of quasiparticles may become non-negligible compared to dielectric loss associated with two-level systems (TLS) in the amorphous surface oxide layers. These findings suggest that the control of oxygen impurities in Nb and Ta-encapsulated Nb thin films during film deposition and qubit fabrication is essential once decoherence from dielectric losses has been substantially mitigated.

We also emphasize that oxygen segregation at GBs can be more problematic for Nb thin films superconducting radiofrequency (SRF) cavity for accelerator applications. Oxygen segregated GBs may have reduced $T_c$ and provide pathways for flux penetration that lead to degradation of quality (Q) factor of Nb thin film SRF cavities [19, 38, 47, 48]. Nb thin film SRF cavities suffer from degradation of Q-factor at high magnetic field compared to bulk Nb SRF cavities and microscopic defects in Nb thin films, including oxygen segregation at GBs and structural changes, could provide potential sources of Q-slope in Nb thin film SRF cavities [46, 49]. The current study suggests that reducing oxygen impurities in Nb thin film and mitigating aging effects of oxygen impurity in Nb can reduce oxygen segregation at GBs and, in turn, may help improve the performance of Nb thin film SRF cavities.

## 5. Conclusions

In summary, we report on atomic-scale analyses of oxygen distribution and segregation at GBs in Nb and Ta-capped Nb thin films for superconducting qubits employing APT and TEM. We find a strong correlation between oxygen concentration in Nb grain interiors and at GBs: higher oxygen concentration in Nb grain interior leads to higher oxygen concentration at GBs with enrichment factors ($C_{GB}/C_{bulk}$) of 2.7 ± 0.3. This result indicates that the control of oxygen concentration in Nb thin films during film deposition and subsequent fabrication process is important to limit the oxygen segregation enrichment at GBs. Ta-capped Nb thin films show slightly lower enrichment factors (2.3 ± 0.3) while the Ta-capping layer itself exhibits higher oxygen concentration and oxygen GB enrichment factors (3.0 ± 0.3), suggesting that Ta-capping layers can trap oxygen and thus affect in-diffusion and oxygen segregation at GBs in the underlying Nb thin film. Given the reduction of $T_c$ with increasing oxygen concentration in Nb, oxygen impurities and segregation at GBs are expected to contribute to the decoherence of superconducting qubits through excitation of quasiparticles. Overall, this study provides insights and methods to mitigate oxygen impurities in Nb thin films and segregation at GBs, thereby providing a pathway for improving the superconducting properties of Nb thin films for quantum devices and superconducting radio frequency (SRF) cavities.

**Acknowledgment**

This work was supported by the U.S. Department of Energy, Office of Science, National Quantum Information Science Research Centers, Superconducting Quantum Materials and Systems Center (SQMS), under Contract No. 89243024CSC000002. Fermilab is operated by Fermi Forward Discovery Group, LLC under Contract No. 89243024CSC000002 with the U.S. Department of Energy, Office of Science, Office of High Energy Physics. We thank Drs. Nathan Sitaraman and Matthias Liepe in Cornell University for the discussion on the effects of oxygen on superconducting properties of Nb. This work made use of the EPIC facility of Northwestern University's NUANCE Center, which has received support from the SHyNE Resource (NSF ECCS-2025633), the IIN, and Northwestern's MRSEC program (NSF DMR-2308691). Atom-probe tomography was performed at the Northwestern University Center for Atom-Probe Tomography (NUCAPT, RRID: SCR_017770). The LEAP tomograph at NUCAPT was purchased and upgraded with grants from the NSF-MRI (DMR-0420532) and ONR-DURIP (N00014-0400798, N00014-0610539, N00014-0910781, N00014-1712870) programs. NUCAPT has also received support from the MRSEC program (NSF DMR-2308691) at the Materials Research Center, the SHyNE Resource (NSF ECCS-2025633), and the Paula M. Trienens Institute for Sustainability and Energy at Northwestern University. This work made use of the Pritzker Nanofabrication Facility, part of the Pritzker School of Molecular Engineering at the University of Chicago, which receives support from Soft and Hybrid Nanotechnology Experimental (SHyNE) Resource (NSF ECCS-2025633), a node of the National Science Foundation National Nanotechnology Coordinated Infrastructure [RRID: SCR_022955]. Parts of this work made use of Rigetti Fab-1 and measurement facilities.

**Supplementary materials**

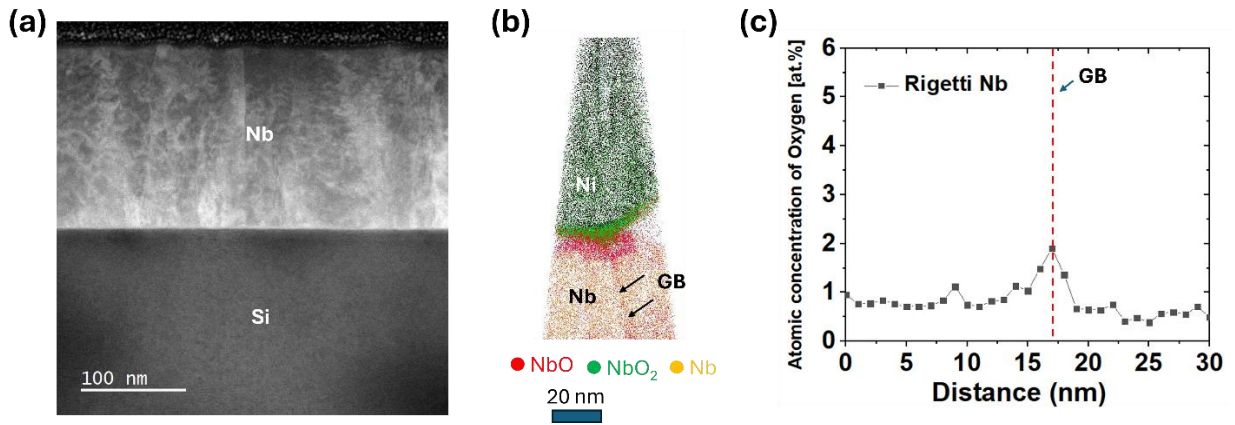

**Fig. S1** (a) ADF-STEM image of Nb thin film on Si (001) substate (Sample 2). It shows typical (110) textured columnar grain structure with 20-50 nm of grain diameter similar to Nb thin film on sapphire (0001) substrate (Sample1) in Fig. 1(a). (b) 3D reconstruction of APT nanotip including a grain boundary (GB) of the polycrystalline Nb thin film (Sample 2). (c) Atomic concentration profile of oxygen across the GB indicates that oxygen concentrations at the GB increase to 1.9 at.% O compared to 0.7 at.% O for Nb grain interiors.

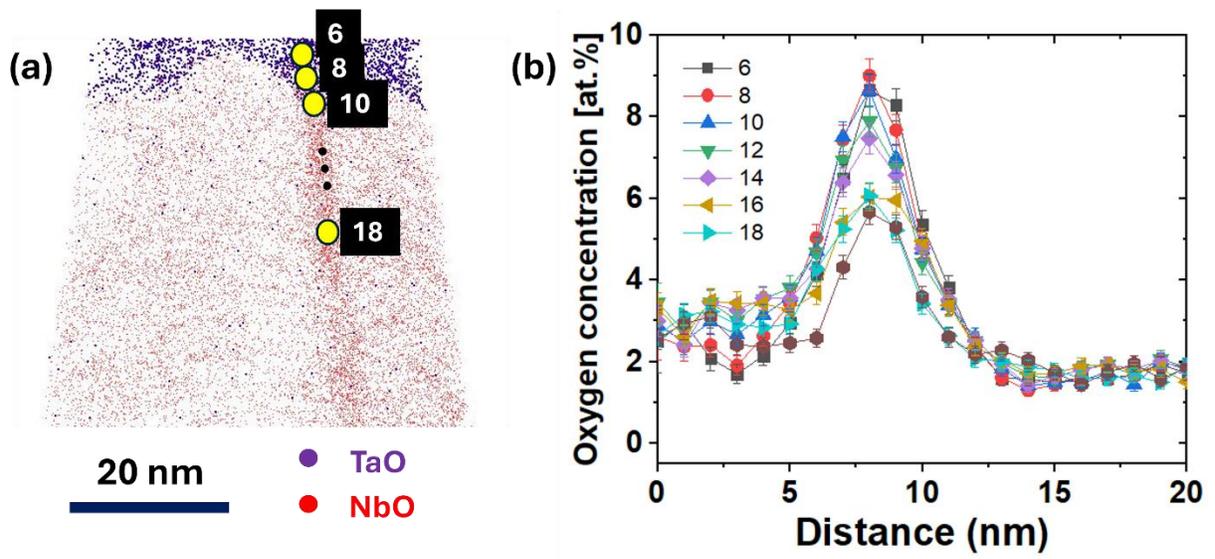

**Fig. S2** (a) Locations where 1-D concentration profiles are collected are denoted as yellow dots in 3-D reconstruction of Ta-O and Nb-O molecular ions in Ta-capped Nb. (b) Overlays of 1D concentration profiles of oxygen across GB in Ta-capped Nb.

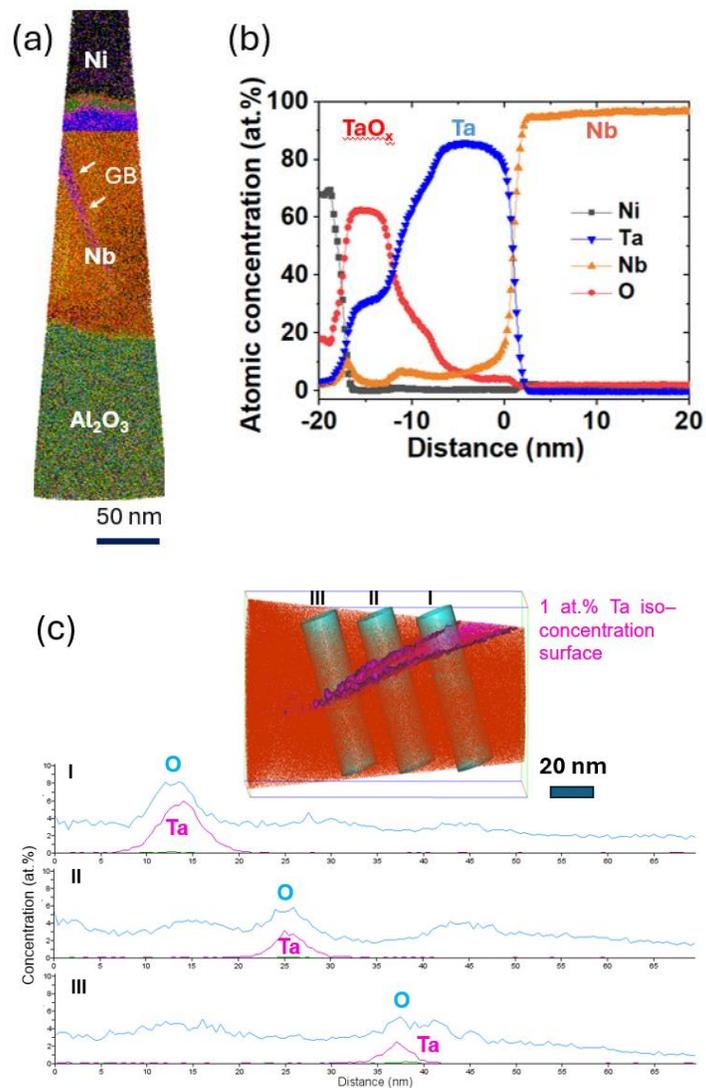

**Fig. S3** (a) 3D reconstruction of APT nanotip for Ta/Nb on Al$_2$O$_3$ (0001) substrate (Sample 3). Ta/Nb films are deposited at RT and annealed at 800 °C for 3 hrs in high vacuum. (b) Atomic concentration profile of Ta/Nb across the Ta capping layer. (c) Atomic concentration profiles across a GB in the Nb film. Oxygen segregation along the GB is observed.